\documentclass{sf2a-conf2017}
\usepackage{graphicx}
\usepackage{hyperref}
\usepackage[]{natbib}  
\usepackage{epstopdf}
\usepackage{longtable}
\usepackage[utf8]{inputenc}

\def\BibTeX{{\rm B\kern-.05em{\sc i\kern-.025em b}\kern-.08em
    T\kern-.1667em\lower.7ex\hbox{E}\kern-.125emX}}
\bibpunct{(}{)}{;}{a}{}{,}  
\newcommand{\kms}{{\mathrm{km~s^{-1}}}}

\newcommand{\teff}{{\mathrm{T_{eff}}}}
\newcommand{\logg}{{\mathrm{\log g}}}

\begin{document}

\TitreGlobal{SF2A 2017}


\title{The flat bottomed lines of Vega}

\runningtitle{HR 7001}

\author{R.Monier}\address{LESIA, UMR 8109, Observatoire de Paris Meudon, Place J.Janssen, Meudon, France}
\author{M.Gebran}\address{Department of Physics and Astronomy, Notre Dame University - Louaize, PO Box 72, Zouk Mikael, Lebanon}
\author{F.Royer}\address{GEPI,UMR 8111, Observatoire de Paris Meudon, Place J.Janssen, Meudon, France}
\author{T. K{\i}l{\i}co\u{g}lu}\address{Department of Astronomy and Space Sciences, Faculty of Science, Ankara University, 06100, Ankara, Turkey}


\setcounter{page}{237}


\maketitle


\begin{abstract}
Using one high dispersion high quality spectrum of Vega (HR7001, A0V) obtained with the \'echelle spectrograph SOPHIE at Observatoire de Haute Provence, we have measured the
centroids of 149 flat-bottomed lines.
The model atmosphere and spectrum synthesis modeling of the spectrum of Vega allows us to provide identifications for all these lines. Most of these lines are due to 
C~I, O~I, Mg~I, Al~I, Ca~I, Sc~II,Ti~II, Cr~I, Cr~II, Mn~I, Fe~I, Fe~II, Sr~II, Ba~II, the large majority being due to neutral species, in particular Fe~I.

\end{abstract}



\section{Introduction}
Vega (HR 7001), the standard A0V spectral type, is one of the 47 northern slowly rotating early-A stars studied by \cite{Royer14}. 
 The low projected rotational velocity of HR 7001, about 24 $\kms$ is due to the very low inclination angle ($i \simeq 0$) while the equatorial velocity $v_{e} \simeq 245 \kms$ is very large \citep{Gulliver94}. Hence Vega is a fast rotator seen nearly pole-on whose limb almost coincides with the equator of the star. At the equator, the centrifugal forece reduces the effective surface gravity which alters the ionization balance and strengthens the local $I_{\lambda}$ profile of certain species.
For these species, the distribution of the Doppler shift is bimodal, ie. arises from the two equatorial regions near the limb.
We have measured all the centroids of all 149 flat-bottomed lines we could find in the high resolution SOPHIE spectrum of Vega.
 We have synthesized all lines expected to be present in the SOPHIE spectrum of HR 7001 in the range 3900 up to 6800 \AA\ using model atmospheres and spectrum synthesis
and an appropriate chemical composition for Vega as derived by \cite{Castelli94}.
The synthetic spectrum has been adjusted adjusted to the SOPHIE spectrum of HR 7001 in order to identify the flat-bottomed lines of HR 7001 
  
\section{Observations and reduction}

A search of the SOPHIE archive reveals that HR 7001 has been observed 78 times at the Observatoire de Haute Provence using SOPHIE from 03 August 2006 to 06 August 2012.
We have used one high resolution (R = 75000) 30 seconds exposure secured with a $\frac{S}{N}$ ratio of about 824 at 5000 \AA\ to search for the flat-bottomed lines. 

\section{Model atmospheres and spectrum synthesis }

The effective temperature and surface gravity of HR 7001 were first evaluated using Napiwotzky et al's (1993) UVBYBETA calibration of Stromgren's photometry.
The found effective temperature $\teff$ is 9550 $\pm$ 200 K and the surface gravity $\logg$ is  $3.98 \pm$ 0.25 dex. This temperature is in very good agreement with the fundamental temperature derived by \cite{Code1976} from the integrated flux and the angular diameter and with the mean temperature and surface gravity derived by \cite{Hill2010}.

A plane parallel model atmosphere assuming radiative equilibrium, hydrostatic equilibrium and local thermodynamical equilibrium was then computed using the ATLAS9 code \citep{Kurucz92}, specifically the linux version using the new ODFs maintained by F. Castelli on her website\footnote{http://www.oact.inaf.it/castelli/}. The linelist was built starting from Kurucz's (1992) gfhyperall.dat file  \footnote{http://kurucz.harvard.edu/linelists/} which includes hyperfine splitting levels.
This first linelist was then upgraded using the NIST Atomic Spectra Database 
\footnote{http://physics.nist.gov/cgi-bin/AtData/linesform} and the VALD database operated at Uppsala University \citep{kupka2000}\footnote{http://vald.astro.uu.se/~vald/php/vald.php}.
A grid of synthetic spectra was then computed with a modified version of SYNSPEC49 \citep{Hubeny92,Hubeny95} to model the lines. The synthetic spectrum was then convolved with a gaussian instrumental profile and a parabolic rotation profile using the routine ROTIN3 provided along with SYNSPEC49.
We adopted a projected apparent rotational velocity $v_{e} \sin i =  24.5 $ km.s$^{-1}$ and a radial velocity $v_{rad} = -13.80 $ km.s$^{-1}$ from \cite{Royer14}.

\section{Determination of the microturbulent velocity}

\begin{figure}[h!]
 \centering
 \includegraphics[width=0.7\textwidth]{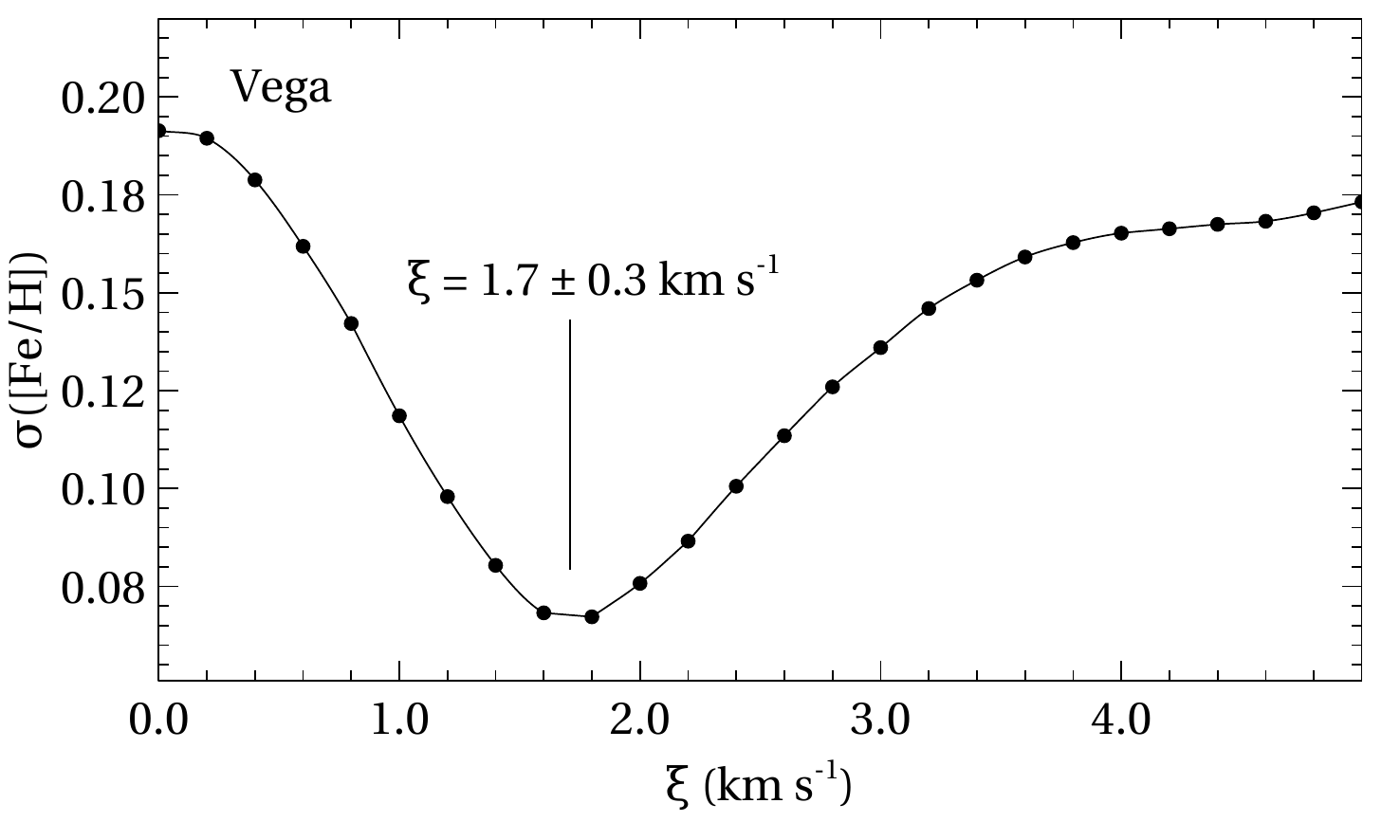}     
  \caption{The derived microturbulent velocity for HR 7001}
  \label{fig1}
\end{figure}

In order to derive the microturbulent velocity of HR 7001, we have derived the iron abundance
[Fe/H] by using 36 unblended Fe II lines for a set of microturbulent
velocities ranging from 0.0 to 2.5 $\kms$. Figure \,\ref{fig1} shows the standard
deviation of the derived [Fe/H] as a function of the microturbulent velocity.
The adopted microturbulent velocity is the value which minimizes the standard deviation ie. for that value, all Fe II lines yield
the same iron abundance, which is [Fe/H] = -0.60 $\pm$ 0.07 dex. Hence iron is found to be underabundant in HR 7001 in agreement with previous abundance determinations 
\citep{Castelli94}. 
We therefore adopt a microturbulent velocity $\xi_t$ = 1.70 $\pm$ 0.04 $\kms$ constant with depth  for HR 7001.

\begin{figure}[h!]
 \centering
 \includegraphics[width=0.7\textwidth]{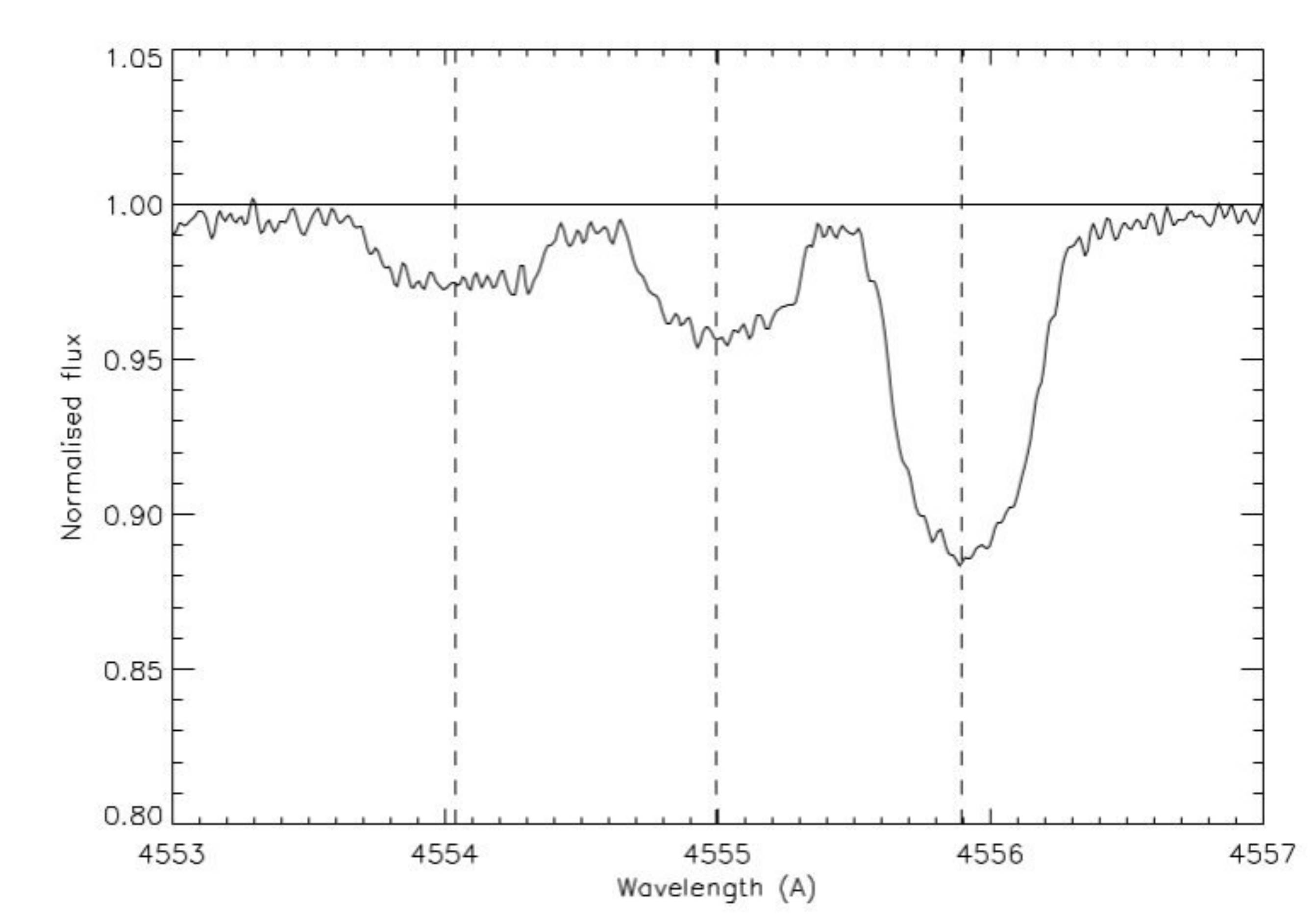}     
  \caption{The flat-bottomed line Ba~II 4554.04 \AA}
  \label{fig2}
\end{figure}

\section{The list of flat-bottomed lines in HR 7001}

An example of flat-bottomed line in the spectrum of HR 7001 is the Ba~II line at 4554.04 \AA\ shown in Fig.~\ref{fig2}. 
Note that the lines of Cr~II at 4554.99 \AA\ and of Fe~II at 4555.99 \AA\ have normal profiles.
All the flat-bottomed lines are collected together with their identifications in Tab.~\ref{longtab-atom}. 
These lines are weak lines due to C~I, O~I, Mg~I, Al~I, Ca~I, Sc~II, Ti~II, Cr~I, Cr~II, Mn~I, Fe~I, Fe~II, Sr~II, Ba~II, the large majority being due to neutral species, in particular Fe~I. Most of the lines we find to be flat-bottomed are also listed in the investigation of weak lines conducted by \cite{Takeda08} in their high signal-to-noise high resolution spectrum of Vega.



\section{Conclusions}

 A systematic search for flat-bottomed lines in the high resolution high quality SOPHIE spectrum of HR 7001 yields 149 lines in the range
3900 \AA\ up to 6800 \AA\ which complete the previous list published by \cite{Takeda08}. Most of these lines are due to 
C~I, O~I, Mg~I, Al~I, Ca~I, Sc~II,Ti~II, Cr~I, Cr~II, Mn~I, Fe~I, Fe~II, Sr~II, Ba~II, the large majority being due to Fe~I.

\begin{longtable}{lllrll}
\caption{Identifications for flat-bottomed lines in Vega}\\
\hline
$\lambda_{\mathrm{obs}}$ (\AA)  & $\lambda_{\mathrm{lab}}$ (\AA) & Species & $\log gf$ & $E_{low}$ & Comments \\ \hline
\endhead
3903.08     & 3902.945   & Fe~I & -0.47 & 12560.933 & \\
3916.45    & 3916.45     &V~II    &-1.060 & 11514.760  & \\ 
3918.64    & 3918.642   &Fe~I   &-0.720  &24338.766  &   \\
3920.31     & 3920.258   & Fe~I & -1.75 & 978.074  &    \\
3922.94    & 3922.912  & Fe~I & -1.65 & 415.933  &     \\
3927.98    & 3927.920     & Fe~I & -1.59  & 888.132  &     \\
3930.31  & 3930.296 & Fe~I & -1.590 & 704.007 &  \\ 
               &3930.304  &Fe~II & -4.030 & 13673.186  &     \\
3932.06   &3932.023  &Ti~II   & -1.780 & 9118.260  & \\
3935.98   & 3935.962& Fe~II & -1.860 & 44915.046 &           \\
3938.32& 3938.289 & Fe~II & -3.890 & 13474.411 &  \\ 
            & 3938.400  & Mg~I & -0.760 & 35051.263  &  \\
3944.03   & 3944.006 & Al~I & -0.620 & 0.000 &  \\
3945.20  & 3945.210   & Fe~II & -4.250 & 13673.186  &   \\
3956.71  & 3956.677  & Fe~I &-0.430 & 21715.731 &     \\
3961.54  & 3961.520  & Al~I &-0.320  & 112.061    &  \\
4002.36  &  4002.483 &  Cr~II  & -2.060  &42897.990   &  ?                    \\
4005.29 & 4005.242 & Fe~I & -0.610 & 12560.933 &  \\ 
4021.88 & 4021.866  & Fe~I & -0.660  & 22249.428  &   \\
4034.49  & 4034.469  & Mn~I  & -0.810  &  0.000   &  \\
               &4034.502  &Mn~I  &-0.810   &  0.000    &     \\
4035.67   & 4035.595 &Fe~I   &-1.100  &34039.315  &   blend \\
               & 4035.627 &V~II   &-0.960   &14461.750   &      \\
	       &4035.694  &Mn~I  &-0.190  &17281.999   &      \\
	       &4035.713  &Mn~I  &-0.190   &17281.999   &       \\
	       &4035.715  &Mn~I  &-0.190   &17281.999    &    \\
4043.99   &4043.897   &Fe~I  &-0.830   &26140.178    & blend       \\
               &4043.977   &Fe~I  &-1.130   & 26140.178   &      \\
	       &4044.012   &Fe~II &-2.410    &44929.549    &       \\
 4057.56 & 4057.461 & Fe~II & -1.550 & 58668.256 &  blend \\
	      &4057.505  &Mg~I &-1.200  &35031.263  &    \\
4068.01  & 4067.978  &Fe~I &-0.430  &25899.986  &     \\
4070.89   &4070.840  &Cr~II &-0.750  &52321.010  &    \\
4072.38   &4072.502  &Fe~I &-1.440  &27666.345 &  \\
4118.57  &4118.545  &Fe~I & 0.280  &28819.952 &  \\
4122.69 & 4122.668 & Fe~II & -3.380 & 20830.553 & \\ 
4132.11 & 4132.058 &Fe~I &-0.650  &12968.554  &    \\
4134.67  &4134.677 &Fe~I &-0.490  &22838.320  &    \\
4143.37  &4143.415  &Fe~I &-0.200  &24574.652  &      \\
4143.86	   &4143.868 &Fe~I &-0.450 & 12560.933  &       \\  
4161.58  & 4161.535  &Ti~II &-2.360  &8744.250   &      \\
4167.34 &4167.271 &Mg~I &-1.600  &35051.263  &   blend \\
              &4167.299 &Fe~II  &-0.560 &90300.626 &      \\
4175.69  & 4175.036 &Fe~I &-0.670 & 22946.815  &      \\
4176.63   &4176.566  & Fe~I &-0.620 & 27166.817   &         \\
4181.74    &4181.755  &Fe~I &-0.180 &22838.320   &     \\
4187.06   &4187.039  &Fe~I &-0.550 &19757.031  &        \\
4187.83   &4187.795  &Fe~I &-0.550  &19562.437  &        \\
4191.49  & 4191.430 &Fe~I &-0.670  &19912.494   &     \\
4198.29  &4198.247  &Fe~I &-0.460   & 27166.817  &  blend     \\
              &4198.304 &Fe~I &-0.720    &19350.891    &       \\
4199.11 & 4199.095 & Fe~I & 0.250 & 24574.655 & \\ 
4202.02  &4202.029 &Fe~I&-0.710  &11976.238   &          \\
4210.40   &4210.343 &Fe~I & -0.870 &20019.633  &          \\  
               &4210.383 &Fe~I &-1.240 &24772.016   &        \\  
4215.60& 4215.519 & Sr~II & -0.140 & 0.000 &  \\ 
4219.40  &4219.360 &Fe~I &0.120  & 28819.952    &       \\
4222.30   &4222.213  &Fe~I &-0.970   &19757.031  &   blend \\
               &4222.381  &Zr~II &-0.900   & 9742.800   &     \\
4226.75  &4226.728  &Ca~I & 0.240  & 0.000     &  \\
4227.45  &4227.427  &Fe~I & 0.230  & 26874.546 &            \\
4236.00  &4235.936  &Fe~I &-0.340  &19562.437  &             \\
4238.80  &4238.810  &Fe~I &-0.280  &27394.689  &      blend          \\
	     &4238.819  &Fe~II &-2.720  &54902.315  &                \\
4250.15  &4250.119  &Fe~I &-0.410  &19912.494  &              \\
4250.80  &4250.787  &Fe~I &-0.710  &12560.933  &     \\
4273.30  &4273.326  &Fe~II &-3.260  &21812.055  &                 \\
4274.80  &4274.797  &Cr~I &-0.230  &0.000      &       \\
4275.60  &4275.606  &Cr~II &-1.710  &31117.390  &      \\
4278.20  &4278.159  &Fe~II &-3.820  &21711.917  &       \\
4282.45 & 4282.403  &Fe~I &-0.810  &17550.180  &   blend    \\
        &4282.490   &Mn~II &-1.680  &44521.521  &                   \\
4284.20   & 4284.188 &Cr~II & -1.860 &  31082.940  &      \\      
4287.90   &4287.872  &Ti~II  &-2.020  & 8710.440     &     \\ 	
4312.90  &4312.864   &Ti~II  &-1.160   &9518.060     &    \\
4325.0 & 4324.999    &Sc~II  &-0.440   & 4802.870    &    \\
4367.6   &4367.659    &Ti~II  &-1.270   & 20891.660   & blend   \\
             &4367.578    &Fe~I  &-1.270  & 24118.816   &     \\
4369.50 &4369.411  &Fe~II &-3.670  &22469.852  &             \\
4371.4  & 4371.367  &C~I  &-2.330   & 61981.822   &       \\
4386.9 &4386.844  &Ti~II &-1.260  &20951.600  &        \\
4394.05 &4394.051  &Ti~II &-1.590  &9850.900   &  \\
4395.08 &4395.033  &Ti~II &-0.660  &8744.250   &       \\
4400.40 & 4400.379 &Sc~II &-0.510  &4883.570  &       \\
4411.10 & 4411.074 &Ti~II &-1.060 & 24561.031  &      \\ 
4417.80  & 4417.719 &Ti~II &-1.430 &9395.710   &     \\
4418.40  &4418.330 &Ti~II &-2.460  &9975.920   &      \\
4450.60  &4450.482  &Ti~II &-1.450  & 8744.250  &    \\
4451.60  &4451.551  &Fe~II & -1.840 &49506.995 &    \\
4454.90  & 4455.027  &Fe~I &-1.090 & 31307.244  &      \\
4464.50  &4464.450  &Ti~II  &-2.080  &9363.620   &     \\
4466.65  &4466.551  &Fe~I &-0.590  &22856.320  &      \\
4473.00  &4472.929  &Fe~II &-3.430  &22939.357   &       \\
4476.10   &4476.019 &Fe~I &-0.570   &22946.815   &  blend  \\
               &4476.076 &Fe~I &-0.290   &29732.735   &      \\
4488.40    &4488.331&Ti~II  &-0.820   &25192.791    &    \\
4494.65   & 4494.563 &Fe~I &-1.140  &17726.928    &     \\
4528.70   &4528.614  &Fe~I &-0.820  &17550.180    &    \\
4529.60   &4529.569  &Fe~II &-3.190  &44929.549     &    \\
4541.60   &4541.524  &Fe~II &-3.050  & 23031.299   &      \\
4554.20  & 4554.033 &Ba~II &         &      0.000       &    15 hfs iso  \\
4582.85  &4582.835    &Fe~II &-3.100  &22939.357     &     \\
4590.0   &4589.958     &Ti~II  &-1.790   &9975.920     &  blend  \\
             &4589.967     &O~I  &-2.390    &86631.153   &      \\
4592.10  &4592.049    &Cr~II &-1.220    & 32854.949   &       \\
4616.60   &4616.629  &Cr~II &-1.290 &32844.760   &      \\
4620.50   &4620.521  &Fe~II &-3.280 &22810.356   &        \\
4666.80   &4666.758   &Fe~II   &-3.330  &22810.356   &        \\
4703.00   &4702.991   &Mg~I &-0.670   &35051.263    &   blend    \\
               &4702.991   &Zr~II  &-0.800    &20080.301    &           \\
4731.50   &4731.453   &Fe~II  &-3.360    &23317.632    &            \\
4736.82    &4736.773  &Fe~I  &-0.740    &25899.986    &   very weak      \\
4775.90    &4775.897  &C~I   &-2.670     &18145.285    &               \\
4780.00   &4779.985   &Ti~II   &-1.370     &16518.860     &           \\
4812.40    &4812.468  &Fe~I    &-5.400      &22249.428   &   \\
4836.20    &4836.229   &Cr~II  &-2.250      &31117.390   &        \\
4890.70    &4890.755   &Fe~I  &-0.430      &23192.497   &      \\
4891.50     &4891.492  &Fe~I   &-0.140      &22996.673   & blend     \\
                 &4891.485  &Cr~II   &-3.040      &31350.901    &      \\
4919.05     &4918.994  &Fe~I  &-0.370      &23110.937    & blend      \\
                 &4918.954  &Fe~I  &-0.340      &33507.120    &         \\
4920.50     &4920.502  &Fe~I  &0.060        &22845.868    &    \\
4932.00      &4932.049  &C~I   &-1.880      &61981.822    &   blend      \\
                  &4932.080  &Fe~II  &-1.730     &83196.488     &	               \\
4934.10      &4934.076  &Ba~II  &-0.150      &0.000           &     \\
4993.40      &4993.565    &N~I  &-2.860     &95475.313     & ?     \\
5004.20      &5004.195    &Fe~II &0.500      &82853.660     &    \\
5052.20      &5052.167    &C~I   &-1.650      &61981.822       &   very flat bottomed  \\
5129.20      &5129.152     &Ti~II    &-1.390    &15257.430         &    \\
5133.70      &5133.688     &Fe~I   &0.140      &33695.394        &          \\
5154.00      &5154.070     &Ti~II    &-1.920      &12628.731       &           \\
5171.60        &5171.596    &Fe~I   &-1.790       &11976.238      &  blend, very extended flat core          \\
                    &5171.640    &Fe~II   &-4.370       &22637.205      &         \\
5185.90         & 5185.913    & Ti~II &  -1.350  &   15265.619  &   \\
5188.70         & 5188.680    & Ti~II & -1.210   & 12758.110 &   \\		    
5192.40         &5192.442   &Fe~II   &-2.020       &5192.442        &         \\
5206.00         &5206.037    &Cr~I   &0.020         & 7593.150      &      \\
5208.40          &5208.425    &Cr~I   &0.160       &7593.150         &   \\
5217.00          &5216.863    &Fe~II   & 0.610      & 84527.779      &  blend   \\
                      &5216.854    &Fe~II   & 0.390      & 84710.686      &     \\    
5226.60          &5226.343     & Ti~II  & -1.300      & 12628.731     &         \\
5226.84          &5226.862    &Fe~I   &-0.550      &24506.914         &    \\
5227.30           &5227.481   &Fe~II   &0.800        &84296.829         & blend \\
                      &5227.323    &Fe~II   &-0.030       &84344.832         &       \\
5237.40           &5237.329    &Cr~II  &-1.150       &32854.311         &   		\\
5255.00          &5254.929     &Fe~II  &-3.230        &26055.422        &      \\  
5266.60          &5266.555      &Fe~I &-0.490        &24180.861           &   \\
5269.60          &5269.537     &Fe~I  &-1.320        & 6928.268             &    \\
5275.01          &5274.964     &Cr~II   &-1.29         &32836.680            &     \\
5291.65         &5291.666     &Fe~II     &0.58          &84527.779             & \\
5313.55         &5313.563      &Cr~II    &-1.650        &32854.949              &   \\
5324.20         &5324.179      &Fe~I    &-0.240        &25899.986              &   \\
5325.50         &5325.503      &Fe~II    &-2.800        &25981.670              &   \\
5329.70         &5329.673      &O~I     & -2.200      & 86627.777              & blend \\
                     & 5329.681    &O~I    & -1.610        & 86627.777              &                   \\
		     & 5329.690    &O~I    & -1.410        & 86627.777             &                 \\
5330.80       &5330.726       &O~I    &   -2.570   & 86631.453              & blend \\
                    &5330.735      &O~I   &-1.710       & 86631.453              &        \\
		    &5330.741     &O~I    &-1.120        &86631.453              &     \\
5336.80         &   5336.771   &  Ti~II & -1.700       &    12758.110     &              \\
5337.60        &5337.732     &Fe~II    &- 3.890          &26055.422               &   \\
                    &5337.772      &Cr~II   &-2.030            &32854.949              &          \\
5362.92        &5362.869        &Fe~II   &-2.740     & 25805.329 &    blend  \\   
                    &5362.957       &Fe~II    &-0.080            &84685.798              &              \\
5367.50       &5367.466       &Fe~I     &0.350            &35611.622              &  \\
5370.00        &5369.961       &Fe~I     &0.350            &35257.323               &   blend        \\
                    &5370.164       &Cr~II     &0.320            &86782.011                &         \\
5371.50        &5371.489       &Fe~I     &-1.650           &7728.059                 &   blend      \\
                    &5371.437       &Fe~I     &-1.240           &35767.561               &         \\
5380.35       & 5380.337       &C~I      & -1.840            & 97770.180              &         \\
5383.30        &5383.369       &Fe~I     &0.500             &34782.420               &      \\
5404.15        &5404.117      &Fe~I       &0.540            &34782.420                & blend \\
                    &5404.151      &Fe~I       &0.520             &35767.561                &     \\
5405.80        &5405.663      &Fe~II       &-0.440            &48708.863                &  blend   \\
                   &5405.775       &Fe~I       &-1.840            &7985.784                  &    \\
5410.90     & 5410.910      & Fe~I        &0.280              & 36079.371               &       \\  
5415.20     & 5415.199      & Fe~I       &0.500               & 35379.205                &        \\
5425.20        &5425.257       &Fe~II       &-3.360             &25805.329              &   \\
5447.00       & 5446.916       & Fe~I      &-1.930             & 7985.784                &     \\
5455.50       & 5455.454        & Fe~I     & 0.300             & 34843.934               &  blend   \\
                   & 5455.609        & Fe~I     & -2.090            & 8154.713                 &       \\
5526.80       &5526.770        &Sc~II        &0.130                 &14261.320                   & \\
5528.40       &5528.405        &Mg~I       &-0.620                &35051.263                   &  \\
5534.80       &5534.847       &Fe~II         &-2.940                 &26170.181                   & blend  \\
                   &5534.890       &Fe~II        &-0.690                 &85048.620                   &  \\
5572.75       &5572.842       &Fe~II       &-0.310                    &27394.689                  &     \\
5586.80       &5586.842       &Cr~II        &0.910                     &88001.361                   & blend   \\
                   &5586.756       &Fe~I       &-0.210                    &27166.817                   &           \\
5588.70      &5588.619       &Cr~II         &-5.550                   &31117.390                   & \\
5615.70       &5615.644        &Fe~I      &-0.140                    &26874.547                    &          \\
5669.0        &5668.943         &C~I       &-2.430                     &68856.328                     &  blend     \\
                  &5669.038         &Sc~II     &-1.120                      &12101.499                      &        \\
 6147.70       &6147.741        &Fe~II       &-2.720                      &31364.440                   &		\\
6149.30       &6149.258        &Fe~II       &-2.720                      &31968.450                   &         \\
6162.0        &6162.173         &Ca~I      &0.100                        & 15315.943        & very weak   \\
6238.35     &6238.392          &Fe~II      &-2.630                       &31364.440                      &    \\
6247.45     &6247.557          &Fe~II      &-2.330                        &31307.949                     &  \\
6417.00     &6416.919       &Fe~II         &-2.740                        &31387.949                  & \\ \hline
\label{longtab-atom}
\end{longtable}

\begin{acknowledgements}
The authors acknowledge use of the SOPHIE archive (\url{http://atlas.obs-hp.fr/sophie/}) at Observatoire de Haute Provence. They have used the NIST Atomic Spectra Database and the VALD database operated at Uppsala University (Kupka et al., 2000) to upgrade atomic data.
\end{acknowledgements}



%
\end{document}